\documentclass[12pt,preprint]{aastex}
\usepackage{color}

\shorttitle{DeFN model for Solar Flare Prediction}
\shortauthors{N. Nishizuka et al.}

\begin{document}

\title{DEEP FLARE NET (DeFN) MODEL FOR SOLAR FLARE PREDICTION}


\author{N. Nishizuka\altaffilmark{1}, K. Sugiura\altaffilmark{2}, Y. Kubo\altaffilmark{1}, M. Den\altaffilmark{1} and M. Ishii\altaffilmark{1}}
\altaffiltext{1}{Applied Electromagnetic Research Institute, National Institute of Information and Communications Technology, 4-2-1, Nukui-Kitamachi, Koganei, Tokyo 184-8795, Japan; nishizuka.naoto@nict.go.jp}
\altaffiltext{2}{Advanced Speech Translation Research and Development Promotion Center, National Institute of Information and Communications Technology}

\begin{abstract}
We developed a solar flare prediction model using a deep neural network (DNN), named Deep Flare Net (DeFN). The model can calculate the probability of flares 
occurring in the following 24 h in each active region, which is used to determine the most likely maximum classes of flares via a binary classification (e.g., $\geq$M class versus 
$<$M class or $\geq$C class versus $<$C class). From 3$\times$10$^5$ observation images taken during 2010-2015 by Solar Dynamic Observatory, we automatically detected 
sunspots and calculated 79 features for each region, to which flare occurrence labels of X-, M-, and C-class were attached. We adopted the features used in Nishizuka et al. 
(2017) and added some features for operational prediction: coronal hot brightening at 131 \,\AA\ (T$\geq$10$^7$ K) and the X-ray and 131 \,\AA\ intensity data 1 and 2 h 
before an image. For operational evaluation, we divided the database into two for training and testing: the dataset in 2010-2014 for training and the one in 2015 for testing. 
The DeFN model consists of deep multilayer neural networks, formed by adapting skip connections and batch normalizations. To statistically predict flares, the DeFN model 
was trained to optimize the skill score, i.e., the true skill statistic (TSS). As a result, we succeeded in predicting flares with TSS=0.80 for $\geq$M-class flares and TSS=0.63 
for $\geq$C-class flares. Note that in usual DNN models, the prediction process is a black box. However, in the DeFN model, the features are manually selected, and it is 
possible to analyze which features are effective for prediction after evaluation. 

\end{abstract}


%
\section{Introduction}

The mechanism of solar flares is a puzzle in solar physics that has remained unsolved for more than one century. They originate from the stored magnetic energy around 
sunspots, emerge from the inner atmosphere, and the impulsive release of the energy produces flares \citep[e.g.,][]{pri02, shi11}. The physical process of sunspot formation and 
flare eruption originating from dynamo action has been intensely studied by observation and theory \citep[e.g.,][]{fle11, mag13, tor13, tak15, tor17}. In particular, the amount 
of solar observation data, which is available in the near real time, has markedly increased. However, it is still difficult to predict flares occurring even within the following 24 h 
by human forecasting.

The occurrence of flares has been studied for a long time. The keys to flare occurrence are the energy storage and triggering processes, which are driven by the emerging flux 
in the photosphere \citep{kus12, ino14, kli14, kan16}. It is empirically known that larger sunspots with a large number of umbra and a more complicated magnetic flux structure 
tend to produce larger flares \citep[e.g.,][]{sam00, lek03, blo12, mcc16}. The energy is globally stored in an active region for 1-2 days, where the amount of energy determines 
the maximum class of flares. When a large class of flares occur, large amounts of magnetic shear and magnetic free energy and the appearance of emerging flux are observed 
along magnetic neutral lines. However, the occurrence of magnetic shear, free energy, and emerging flux does not necessarily foretell a large flare \citep{geo07, mas10, fal14}.

On the other hand, trigger mechanisms of flares are locally observed in a shorter time scale 2-3 h before a flare occurs, as a rapid change in the magnetic field or accumulation 
near magnetic neutral lines \citep{moo04, sai07, nis09, bam13, wan17}. These are sometimes associated with pre-flare events, such as ultraviolet (UV) brightening in the 1600 
\,\AA\ continuum and coronal brightening. UV 1600 \,\AA\ brightening represents upper photospheric heating by a small-scale energy release driven by emerging flux. The 
repetition of flares is frequently observed \citep[e.g.,][]{zir88, zir91, whe04}, and pre-flare brightening in soft X-ray and radio emissions has also been reported \citep{asa06, chi06, chi07, 
sia09}, as well as turbulence and reconnection outflow before the release of impulsive energy \citep{wal10, mck13, har13}.

To deal with the recent large amount of solar observation data, a new approach has been developed using machine-learning algorithms including a neural network \citep{qah07, 
col09, hig11, ahm13}, a regression model \citep{lee07, son09}, a k-nearest-neighbor algorithm \citep[k-NN;][]{li08, hua13, win15, nis17}, a support vector machine \citep[SVM;][]{
qah07, bob15, mur15, alg15, bou15, rab17, sad17}, the least absolute shrinkage and selection operator \citep{ben18, jon18}[LASSO;][], a random forest \citep{liu17}, an extremely randomized trees \citep[ERT;][]{nis17}, an unsupervised 
fuzzy clustering \citep{ben18}, and an ensemble of four prediction models \citep{gue15}. Machine learning can clarify which feature is most effective for predicting flares. However, 
it is still not clear which model is best for prediction in an operational setting i.e., the chronological splitting of the dataset into training and testing datasets.

Nishizuka et al. (2017) compared three machine-learning algorithms for flare prediction: the k-NN, the SVM, and the ERT algorithm. They found that the k-NN shows the best 
performance in the case of random shuffling and dividing the dataset. However, it was also found that the performance of the models changes with differences in the splitting of 
the dataset into training and testing datasets. In the case of random shuffling and dividing, samples are separated into training and testing datasets within 24 h. Because these 
datasets, especially those of magnetograms, are similar to each other, the simple criterion of the distance between two data can give us better prediction, which is an advantage 
of the k-NN algorithm.

In contrast, chronological splitting of the dataset into training and testing datasets makes the prediction of flares more difficult, because the training and testing datasets become 
completely separated. Because we found that our machine-learning models in Nishizuka et al. (2017) using k-NN, SVM, and ERT cannot predict flares with significant precision in 
an operational setting, we focused on the DNN \citep{hin06, lec15} algorithms, which can generally maximize the prediction accuracy or minimize the cost function. Thus, we 
developed a flare prediction model using the DNN, named the Deep Flare Net (DeFN) model. In the training datasets, we included new features related to the trend of data, i.e., the 
maximum intensities of soft X-ray and EUV 131 \,\AA\ emissions 1 and 2 h before an image. In section 2, we briefly explain neural networks. In section 3 we give an overview of 
our prediction model, which is explained in detail in section 4. The prediction results are described in section 5 and a discussion and conclusion are given in section 6.

\clearpage
%
%
\section{Basic Architecture of Neural Networks}   

A neural network (NN) is a set of linear and nonlinear conversions of input data. Nonlinear conversion represented by an activation function enables representations 
that cannot be made by linear conversions, for example, the consideration of curves as the separators or the distortion/curvature of the space of data. In other word, a NN is a 
classifier consisting of several layers, which repeat linear (affine) and nonlinear conversions to search for the most suitable mapping of the original dataset into a higher dimension 
to be linearly separated. One layer converts the input ${\bf x}$ into the output ${\bf y}$ as follows, 
\begin{equation}
{\bf y} = f({\bf x}; W, {\bf b}) = f(W{\bf x} + {\bf b}).
\end{equation}
Here, $W$ is a weight matrix and ${\bf b}$ is bias. This conversion is represented by an image of Figure 1. The function $f$($\cdot$) is called the activation function in the field of 
machine-learning \citep{hah00, bis06, glo11}, whereas its inverse function $f^{-1}$ is the link function in the statistical field. We determine parameters in Table 1. 

As the activation function for the first to the second last layers, we adopted the rectified linear function or rectified linear units \citep[$ReLU$;][]{nai10}, 
\begin{equation}
ReLU(x) = \log (1 + \exp (x))  \simeq max(x, 0)
\end{equation}
This function is commonly used as an activation function giving a value greater than zero, resulting in a sparse distribution and a faster calculation because the 
derivative is one. To give a binary classification of flares (e.g., $\geq$M class versus $<$M class or $\geq$C class versus $<$C class), we calculate the probability of the two classes 
as the output. For this purpose, we used a softmax function or a normalized exponential function for the last layer,
\begin{equation}
Softmax( x_i ) = \frac{ \exp (x_i) }{ \sum_{j=1}^N \exp (x_j) }
\end{equation}

Figure 1(a) shows a normal NN with one hidden layer and Figure 1(b) shows a simplified representation of Figure 1(a). In Figure 1(b), the hidden layer is still represented 
by a square, but the input and output layers are simplified to $x$ and $y$, respectively. The biases of the nodes, $b_1$ and $b_2$, are omitted. The arrows indicate multiple connections 
between each unit through linear conversion. The description of the DeFN model later in Figure 4 follows this simplified style.

\begin{figure}[hbtp]
\epsscale{.70}
\plotone{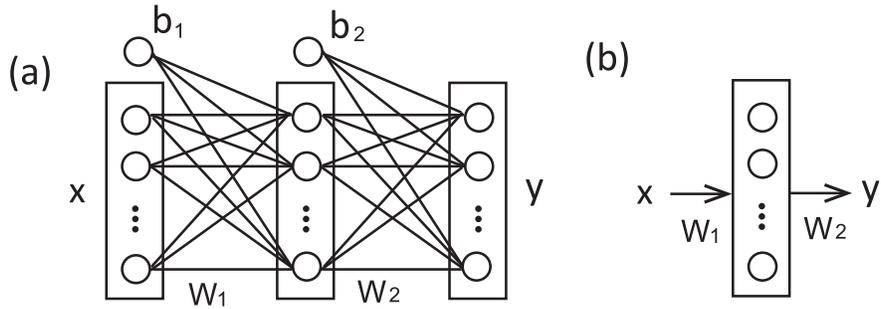}
\caption{Input and output of two layers. $W_1$ and $W_2$ are weight matrices, and $b_1$ and $b_2$ are bias vectors. Many links are complicated and often represented by a more 
simple style at the right hand side. \label{fig1}}
\end{figure}
 
\begin{deluxetable}{llllrl}
\tabletypesize{\normalsize}
\tablecaption{Symbol notations. \label{tbl1}}
\tablewidth{0pt}
\startdata
\tableline
\colhead{x, {\bf x}} & \colhead{Arbitrary input parameters}                                                                                         \\[+0.1cm]
\colhead{y, {\bf y}} & \colhead{Arbitrary output parameters (discrete or continuous)}                                                    \\[+0.1cm]
\colhead{N} & \colhead{The number of training samples}                                                                                        \\[+0.1cm]
\colhead{K} & \colhead{The number of classes/categories}                                                                                    \\[+0.1cm]
\colhead{${\bf y}_k^*$ = $\{$ $y_{nk}^*|$ $n$=1,....,N $\}$ } & \colhead{Correct label of n-th training sample}             \\[+0.1cm]
\colhead{$p$(${\bf y}_k$) = $\{$ $p$($y_{nk}$)$|$ $n$=1,....,N $\}$ } & \colhead{Estimated value of probability of ${\bf y}_k$} \\[+0.1cm]
\enddata
\end{deluxetable}

%
%
\section{Overview of Deep Flare Net (DeFN)}

We introduce the procedures of our flare prediction model as follows (see also Fig. 2). (i) First, observation data are downloaded from the web archives of Solar Dynamic Observatory 
\citep[SDO;][]{pes12} and the Geostationary Operational Environmental Satellite (GOES), such as the line-of-sight magnetogram, vector magnetogram, 1600 \,\AA\ and 131 \,\AA\ 
filter images, and the light curves of the soft X-ray emission. (ii) Second, active regions (ARs) are detected from full-disk images of the line-of-sight magnetogram, and the ARs are 
tracked using their time evolution. (iii) For each AR, features are calculated from multiple wavelength observations, and flare labels are attached to the solar feature database if an 
X/M/C-class flare occurs within 24 h after an image. (iv) Supervised machine learning by a DNN is carried out with a 1 h cadence to predict the maximum class of flares occurring 
in the following 24 h because the features are extracted at a cadence of 1 h. When the corresponding data was missing, the nearest data within 30 min was first searched 
for, and if this search failed, the prediction was skipped. Our observation data are from June 2010 to December 2015, which were taken by SDO, launched in February 2010. During 
this period, 26 X-class, 383 M-class, and 4054 C-class flares were observed on the disk, accounting for 90\% of the flares observed during the period. 

We used the line-of-sight magnetogram taken by the Helioseismic and Magnetic Imager \citep[HMI;][]{sche12, scho12} on board SDO, as well as the vector magnetogram. The UV 
continuum of the upper photosphere and the transition region were taken by the 1600 \,\AA\ filter of the Atmospheric Imaging Assembly \citep[AIA;][]{lem12} on board SDO, and 
the hot coronal brightening in the flaring region was taken by the 131 \,\AA\ filter on board SDO. The full-disk integrated X-ray emission over the range of 1-8 \,\AA\ was observed 
by GOES. The time cadence of the line-of-sight magnetogram is 45 s, that of the vector magnetogram is 12 min, those of the 1600 \,\AA\ and 131 \,\AA\ filters are both 12 s, 
and that of GOES is less than 1 min. Thus, the total size of the observation dataset is so large that we reduced the cadence to 1 h, in accordance with the forecast operation 
every hour from 00:00 UT.

\begin{figure}[hbtp]
\epsscale{.85}
\plotone{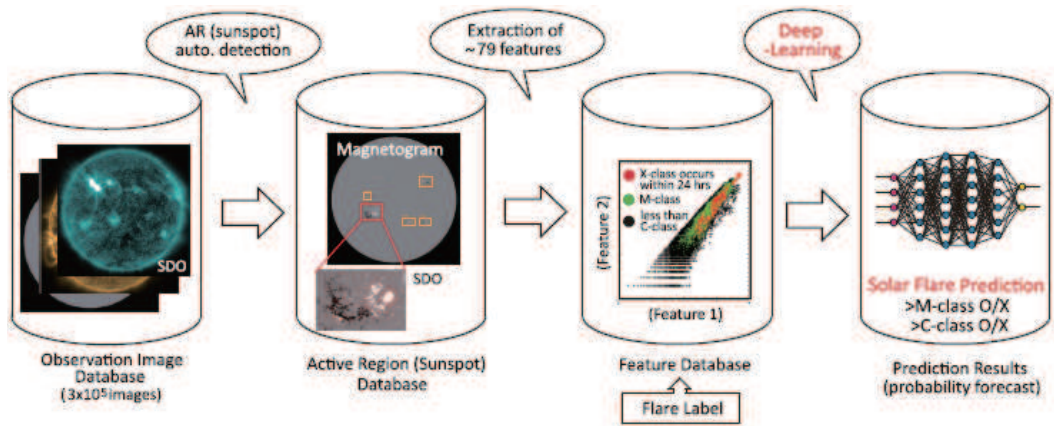}
\caption{Flow chart of our DeFN model of solar flare prediction. \label{fig2}}
\end{figure}

\clearpage
%
%
\section{Details of Deep Flare Net (DeFN) model}
\subsection{Detection of ARs}

First, we detected ARs to extract solar features from the images of the downloaded observation database. We used 3$\times$10$^5$ full-disk images of the line-of-sight magnetogram 
for detection with a reduced cadence of 1 hr. The line-of-sight magnetogram was selected for AR detection because it is less noisy than the vector magnetogram and more suitable for 
the processing carried out for detection. After determining the ARs in magnetogram images using a threshold of 140 G, the frame coordinates of the ARs were applied to other images 
with different wavelengths (Fig. 3). The detection rules are the same as in Nishizuka et al. (2017). We neglected ARs whose frames are across the limb determined by a 
threshold on intensity in photospheric images. By tracking the same ARs, we numbered them for identification; these numbers are different from the NOAA AR numbers. 

\begin{figure}[hbtp]
\epsscale{.55}
\plotone{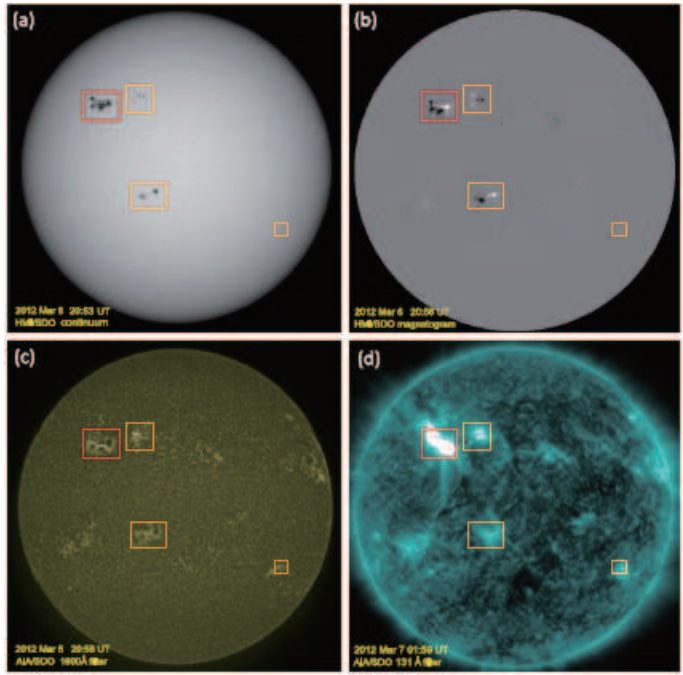}
\caption{Full-disk images of (a) the while light intensity taken by HMI/SDO with detected active regions framed in yellow or red, (b) the line-of-sight magnetogram 
taken by HMI/SDO, (c) the UV continuum taken with the 1600 \,\AA\ filter of AIA/SDO, and (d) the coronal brightening taken by the 131 \,\AA\ filter of AIA/SDO. 
The region with a red frame produced X5.4 flare, 3 hr after this image was taken. \label{fig3}}
\end{figure}

\subsection{Extraction of solar features}

Using the database of detected ARs, we next extracted solar features from each AR. We adopted solar features used in previous papers, which were extracted from the line-of-sight 
magnetogram \citep[e.g.,][]{ste11, ahm13}, the vector magnetogram \citep{lek03, bob15}, the UV continuum taken by the 1600 \,\AA\ filter \citep{nis17}, and GOES X-ray data 
in the range of 1-8 \AA. These extracted features are the same as Table 1 in Nishizuka et al. (2017). Furthermore, in this study, we extracted the feature of hot 
coronal brightening observed by the 131 \,\AA\ filter of AIA/SDO, showing the emissions of iron-20 (Fe$_{XX}$) and iron-23 (Fe$_{XXIII}$) at temperatures greater than 10$^7$ K, 
for the first time. We also added the data of 131 \,\AA\ and GOES X-ray emissions 1 and 2 h before an image, because they are expected to be useful for the operational prediction. 
Since the data of the GOES X-ray is integrated over the full disk, the individual X-ray intensity is not available for each AR, but the integrated value is used. Instead of the individual 
X-ray intensity, we calculated the maximum intensity of 131 \,\AA\ emission for each AR, so that every AR has its own dataset of 131 \,\AA\ emission, which is more efficient 
for prediction.

\subsection{Classification by DeFN model}

Figure 4 shows our model, which we name Deep Flare Net (DeFN). This model is based on deep-learning techniques and consists of multilayers. The input is our solar feature database, 
i.e., 79-dimensional vectors of standardized features, and the output is the prediction probability of each class of flares, $p(y)$. Here $y$ = ($y_1$, $y_2$) is the class of flares: ($y_1$, 
$y_2$) = (0, 1) for $\geq$M-class flare events and ($y_1$, $y_2$) = (1, 0) for $<$M-class or non-flare events. We calculate two probabilities for $\geq$M-class flare events and for 
$<$M-class or non-flare events, and finally we simply select the category with the larger probability. 

Each layer of the neural network in Figure 4 represents a map from the input to the output with a linear combination and an activation function (generally nonlinear). In this model, 
we used ReLU \citep{nai10} and softmax function only for the last layer as activation functions. To increase the number of layers, we used a simple skip connection \citep{he15}, 
which has the role of increasing the precision of the model and avoiding the divergence in the case of multiple layers (for more detail, see Appendix A.1). The notation ${\bf BN}$ 
represents batch normalization \citep{iof15}, which standardizes the input parameters of each layer to stabilize the training and to improve the precision (see Appendix A.2 for 
more details).

\newcommand{\argmax}{\mathop{\rm argmax}\limits}

To maximize the prediction accuracy, the training was performed to minimize a cost function $J$. For a classification problem, parameters are optimized to minimize the cross entropy. 
However, since the flare occurrence ratio is imbalanced, we optimized parameters instead by the summation of the weighted cross entropy, 
\begin{equation}
J = \sum_{n=1}^N \sum_{k=1}^K w_k y_{nk}^* \log p(y_{nk}).
\end{equation}
The parameters used here are summarized in Tables 1 and 2. In Table 2, we used the recommended default values of Adam\footnote{Adam (Adaptive moment estimation) 
is a method for stochastic optimization, which is extended from AdaGrad, RMSprop, and AdaDelta \citep{kin14}.}. $w_k$ is the weight of each class, which is the inverse of the class 
occurrence ratio. The number of nodes and the batch size were investigated in the range of 50-200. The architecture of the DeFN model with 5-9 layers was surveyed 
by attaching or detaching skip connections.

The output of the flare prediction for a dataset is determined by the value of $k$ that gives the maximum probability: 
\begin{equation}
\hat{y} = \argmax_{k} p(y_k).
\end{equation}
In the two-class classification ($\geq$M-class or $<$M-class), we simply select the category with the larger probability.
   
\begin{figure}[hbtp]
\epsscale{.90}
\plotone{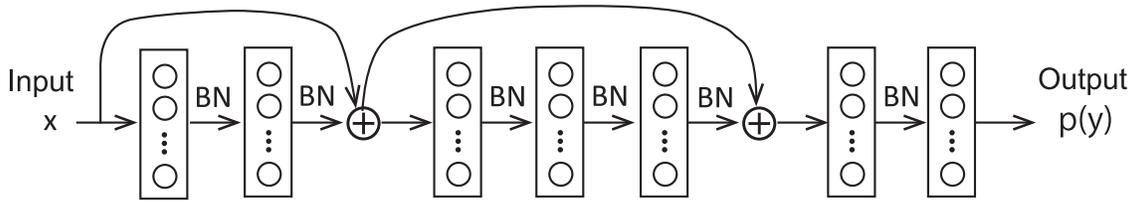}
\caption{Overview of Deep Flare Net (DeFN) model.\label{fig4}}
\end{figure}

\begin{deluxetable}{llllrl}
\tabletypesize{\normalsize}
\tablecaption{Parameter settings of the DeFN model. \label{tbl2}}
\tablewidth{0pt}
\startdata
\tableline
\colhead{Optimization method} & \colhead{Adam (Learning rate = 0.001, $\beta_1$ = 0.9, $\beta_2$ = 0.999)}   \\[+0.1cm]
\colhead{$w_k$ =} & \colhead{(1, 50) for $\geq$M-class flares, (1, 4) for $\geq$C-class flares}               \\[+0.1cm]
\colhead{Num. of nodes =} & \colhead{79 (input), 200, 200, 79, 200, 200, 79, 200, 2 (output)}                    \\[+0.1cm]
\colhead{Batch size =} & \colhead{150}                                                                                               \\[+0.1cm]
\enddata
\end{deluxetable}

\clearpage
%
%
\section{Results}

Using the DeFN model and solar feature database, we performed solar flare prediction within the following 24 h in an operational setting. We divided the database of 2010-2015 into two: 
the dataset in 2010-2014 for training and validation, and the one in 2015 for testing (Fig. 5). These chronological datasets for training and testing make it more difficult 
to predicting flares than the randomly shuffled and divided datasets \citep[e.g.,][]{bob15, mur15, nis17, nis18}. Using mini-batches of the training datasets, the optimization 
was repeated many times. Here a mini-batch corresponds to the number of training samples in one forward and backward pass \citep{goo16}. A mini-batch is randomly selected from all 
the training samples, avoiding overlapping. The update of the weight parameters in an epoch, i.e., one forward and backward pass, is stabilized to converge faster by using mini-batches.

We performed two-category predictions: (i) $\geq$M-class flare events or $<$M-class/non-flare events, (ii) $\geq$C-class flare events or $<$C-class/non-flare events. The number 
of X-class flares was small during 2010-2015; thus, X-class flares were not solely predicted by the DeFN model because the training sample was insufficient. We evaluated 
models using test datasets in each epoch, i.e., one forward pass and one backward pass of all the training examples, and we selected the model giving the maximum skill score, named 
the true skill statistic (TSS).

The prediction results obtained by the DeFN model are summarized in Figure 5. The contingency tables show the number of true positive (TP), true negative (TN), false positive 
(FP), and true negative (TN) events. TN is very large because flare prediction is an imbalanced problem. TP for $\geq$C-class flares is much larger than for $\geq$M-class flares. 
This is a result of the underlying distribution of the data (more flares for $\geq$C than $\geq$M). The occurrence of flares is over predicted, resulting in a large FP. 
Here we note that TSS is defined by two terms: $TSS$ = TP/(TP+FN) - FP/ (FP+TN). Thus, if we over predict the flare occurrence, the first term increases and the number of missing 
flares decreases. At the same time, the second term increases, but TN is so large that the second term changes only slightly. Therefore, the net value of TSS tends to increase when 
over predicting flares.

From the contingency tables, we calculated six skill scores: the probability of detection (POD), the critical success of index (CSI), the false alarm ratio (FAR), Heidke skill score (HSS), 
TSS, and accuracy (for the determination of the skill scores, see Appendix). We show the results in Table 3. We achieved TSS=0.80 for $\geq$M-class flare prediction in an operational 
setting and TSS=0.63 for $\geq$C-class flare prediction in an operational setting. These results are better than human forecasts \citep{cro12, dev14, kub17} and other baseline 
algorithms such as SVM, kNN and ERT (see Table 3).

\begin{figure}[hbtp]
\epsscale{.75}
\plotone{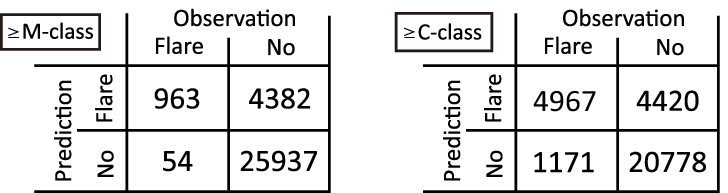}
\caption{Prediction results of $\geq$M-class and $\geq$C-class flares occurring in the following 24 hours obtained by DeFN model for the operational chronological 
training/testing datasets. \label{fig5}}
\end{figure}

\begin{deluxetable}{lll|lrl}
\tabletypesize{\normalsize}
\tablecaption{Skill scores of prediction results in an operational setting (the dataset in 2010-2014 for training, and the dataset in 2015 for testing), by DeFN 
and other baseline algorithms such as SVM, kNN, and ERT with default parameter settings: Probability of Detection (POD), Critical Success of Index (CSI), False Alarm Ratio 
(FAR), Heidke skill score (HSS), True skill statistic (TSS), and accuracy. \label{tbl3}}
\tablewidth{0pt}
\startdata
\tableline
& \multicolumn{2}{c}{DeFN} & \colhead{SVM} & \colhead{kNN} & \colhead{ERT} \\
\colhead{} & \colhead{$\geq$M-class} & \colhead{$\geq$C-class} & \colhead{$\geq$M-class} & \colhead{$\geq$M-class} & \colhead{$\geq$M-class} \\[+0.1cm]
\hline
\colhead{POD} & \colhead{0.95} & \colhead{0.81}  & \colhead{0.36} & \colhead{0.19} & \colhead{0.17}      \\[+0.1cm]
\colhead{CSI}  & \colhead{0.18} & \colhead{0.47}  & \colhead{0.16} & \colhead{0.16} & \colhead{0.16}      \\[+0.1cm]
\colhead{FAR} & \colhead{0.82} & \colhead{0.47}  & \colhead{0.75} & \colhead{0.53} & \colhead{0.36}      \\[+0.1cm]
\colhead{HSS} & \colhead{0.26} & \colhead{0.53}  & \colhead{0.27} & \colhead{0.26} & \colhead{0.26}      \\[+0.1cm]
\colhead{TSS} & \colhead{0.80} & \colhead{0.63}  & \colhead{0.33} & \colhead{0.19} & \colhead{0.17}      \\[+0.1cm]
\colhead{Accuracy} & \colhead{0.86} & \colhead{0.82} & \colhead{0.94} & \colhead{0.97} & \colhead{0.97}  \\[+0.1cm]
\enddata
\end{deluxetable}

\clearpage
%
\section{Summary and Discussion}
 
We developed a flare prediction model, named Deep Flare Net (DeFN), with supervised machine-learning techniques, particularly a DNN. We used solar observation images of a vector 
magnetogram, UV 1600 \,\AA\ brightening in the photosphere, and coronal brightening in soft X-ray and EUV 131 \,\AA\ emissions. By detecting ARs as shown in Nishizuka et al. 
(2017), we extracted novel features, i.e., the features added in previous work, namely, the histories of GOES X-ray and AIA 131 \,\AA\ emission intensities. Then, we attached 
flare labels to the feature database. We divided the dataset into two: the dataset in 2010-2014 for training and the dataset in 2015 for testing. Then, we ran the DeFN model to 
predict the maximum classes of flares that occur in the following 24 h after observation images by calculating the probabilities of flares in each region with a binary classification 
(i.e., $\geq$M-class versus $<$M-class or $\geq$C-class versus $<$C-class; Fig. 6). 

In an operational setting, our DeFN model achieved skill scores of TSS =0.8 for $\geq$M-class flares and TSS=0.63 for $\geq$C-class flares. The number of X-class flares in the 
testing dataset was insufficient for a DNN; thus, the results of X-class flare prediction are not shown here. Generally speaking, a DNN is constructed to maximize the accuracy of 
prediction, making it highly suitable for flare prediction. In fact, we achieved good skill scores using fully shuffled and divided datasets with other machine-learning methods, such as 
SVM, k-NN, and ERT, in Nishizuka et al. (2017). However, we found that in an operational setting, the models do not have the same performance, because the operational setting 
is more difficult than the setting with the shuffled and divided datasets (see Table 3).

Note that in DNN models, cross-validation is not used. Instead, parameters are updated every epoch in the DNN models. An epoch corresponds to one forward/backward 
pass of all the training examples. The parameters in the first epoch, second epoch, and n-th epoch are different. In each epoch, the test dataset is evaluated using the model with 
updated parameters. Through the iterations in epochs, the cost function decreases and the generalization error, TSS, the updated parameters and the model also change. Finally, 
from all the models, we select the best model with the highest test-set TSS. The easiest evaluation is to only use the test datasets, while, for greater precision, both test and 
validation datasets are prepared for the evaluation. In this study, to deal with the sample datasets effectively, we selected the best model only using the test datasets.

The advantage of this DeFN model is that the features are manually selected and they can be analyzed to search for the most effective features for flare prediction. This is different 
from other DNN models, where a convolution network extracts imaging features that humans cannot understand, making it impossible to explain the high precision obtained using a 
feature database \citep{nag17, par17, yi17, had17, hua18}. Using the DeFN model, the ranking of the features can be shown in principle, although this requires huge computational 
resources so is not shown here\footnote{The calculation time using a GPU is estimated as 2-3 h $\times$ 80 features = 160-240 h = 7-10 days. This is 8-36 times 
that of other machine-learning models (e.g., SVM, ERT, kNN) run on a CPU.}. The ranking is not derived from the weights of the features. By removing features one-by-one from 
the original feature database, the ranking of features can be investigated on the basis of the skill score variation.

In the daily forecast operations at NICT Space Weather Forecast Center, which use the knowledge of experts, TSS was 0.21 for X-class flares and 0.50 for $\geq$M-class flares during 
the period 2000-2015 \citep{kub17}. At the Solar Influences Data Center of the Royal Observatory of Belgium, TSS was 0.34 for $\geq$M-class flares during the period 2004-2012 
\citep{dev14}. At NOAA Space Weather Prediction Center, TSS was 0.49 for X-class flares, 0.53 for $\geq$M-class flares, and 0.57 for $\geq$C-class flares \citep{cro12}. The Met 
Office in UK also reported their prediction results \citep{murr17}, although TSS was not derived. Therefore, by comparing TSS our DeFN prediction model appears to achieve better 
performance than human operations. For the verification of forecasting skill, TSS is recommended in the space weather forecasting community \citep{blo12} because it is a 
base-rate-independent measure. However, there is still discussion on which measure is more suitable for verifying rare-event forecasts \citep{bar16}.

\begin{figure}[hbtp]
\epsscale{.55}
\plotone{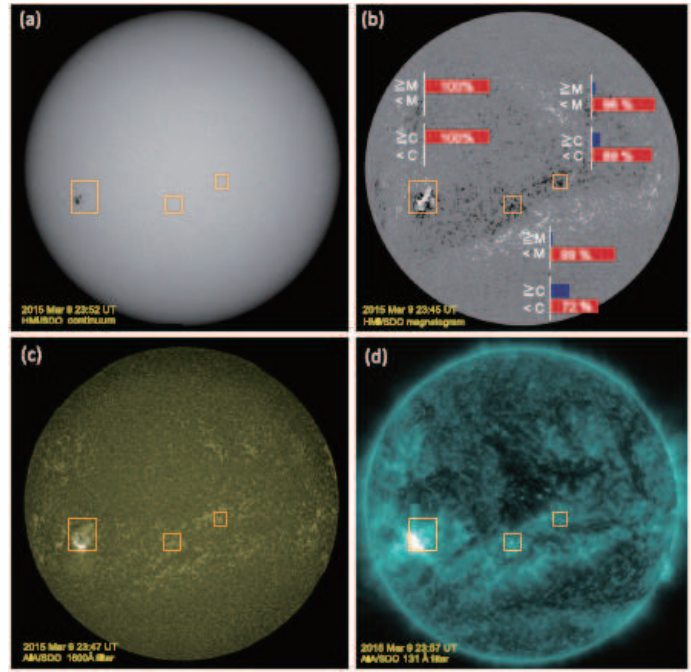}
\caption{Four solar images of SDO observed at different wavelengths, i.e., using white light (HMI), photospheric magnetogram (HMI), 1600, and 131 \,\AA\ filters (AIA). 
The probabilities of predicted flares with a binary classification on each AR are attached to the image of the magnetogram. \label{fig9}}
\end{figure}


\clearpage
%
%
\appendix
\section{Appendix: Details of DeFN model}   
\subsection{Skip Connection (Residual Network)}

DeFN model includes simple skip connections. If we simply increase the number of layers, the precision of the model decreases. To resolve this problem, an algorithm to learn 
the residual function of each layer and to optimize the parameters was developed \citep[residual network;][]{he15}, which enabled to increase the number of 
layers successfully. 

A skip connection (or identity mapping) is illustrated in Figure 8 and modeled as follows:
\begin{equation}
{\bf y} = F({\bf x}; \lambda )+{\bf x},
\end{equation}
where ${\bf x}$ is the input and $F$ is a residual function, which is a conversion map connecting all the layers between the input and the output. $\lambda$ represents all the 
parameters between the input and the output. It is difficult for the derivative of the loss function $\varepsilon$ to become zero. Using ReLU as an activation function, the 
calculation by the gradient descent method does not vanish or explode. The S layers between the input and output of the connection (Fig. 7) can be neglected 
or skipped if they are not worthy of learning. This enables the number of layers to be increased, allowing more complicated models to be represented.

\begin{figure}[hbtp]
\epsscale{.40}
\plotone{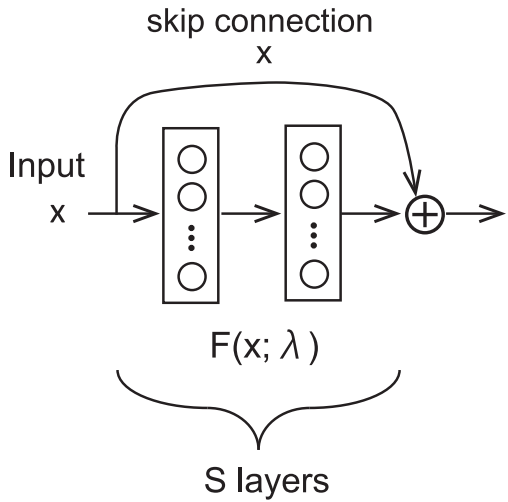}
\caption{The part of a skip connection shown by $\oplus$ and an arrow. \label{fig8}}
\end{figure}

\subsection{Batch Normalization (BN)}

The description $BN$ in Figure 4 represents the batch normalization. The batch normalization standardizes the input parameters at each layer during training and stabilizes the 
training process \citep{iof15}. The output of the batch normalization $\{ y_1 \cdots y_m \}$ is described as follows,
\begin{eqnarray}
x_i = \frac{x_i - \mu}{\sqrt{\sigma_B^2 + \epsilon}}, \\
y_i = \gamma x_i + \beta,
\end{eqnarray}
where $\mu_B$ and $\sigma_B$ are the average and the dispersion of mini-batch, respectively. Here a mini-batch corresponds to the number of training examples 
in one forward/backward pass, and the mini-batch is normalized in batch normalization. $\gamma$ and $\beta$ are parameters optimized in the same way as the weight parameters 
\citep{iof15}. $\epsilon$ is a small value used to avoid dividing the equation by zero. A suffix $i$ is an index of nodes of hidden layers. 

\subsection{Cost Function}

The training is performed to minimize a cost function $J$ or loss function. For classification problems, parameters are usually optimized to minimize the cross entropy. The 
cross entropy between $p$(${\bf y}_k^*$) and $p$(${\bf y}_k$) is determined by the following equation,
\begin{equation}
H_{CE} = \sum_{k=1}^K p({\bf y}_k^*) \log p({\bf y}_k),
\end{equation}
where we omitted n representing the n-th sample, for simplicity. $p$(${\bf y}_k^*$) is the initial probability of correct labels ${\bf 
y}_k^*$, i.e., 1 or 0, while $p$(${\bf y}_k$) are estimated values of probability. The components of ${\bf y}_k^*$ are 1 or 0, and thus, $p$(${\bf y}_k^*$) = 
${\bf y}_k^*$. 

It is stressed here that the flare occurrence rate is imbalanced. In such a case, the summation of the above cross entropy does not lead to a better precision. 
In this model, we adopted the summation of the weighted cross entropy as the cost function as below:
\begin{equation}
J = \sum_{n=1}^N \sum_{k=1}^K w_k y_{nk}^* \log p(y_{nk})
\end{equation}
Here, $w_k$ is weight of each class, and we explicitly include n to indicate n-th sample. The inverse ratio of the class occurrence is sometimes used for $w_k$. 

\clearpage
\section{Appendix: Evaluation Method}
\subsection{Definitions of Skill Scores}

To evaluate our prediction results of flares, we used some well-known skill scores. The probability of detection (POD) or the recall, the critical success 
of index (CSI) or the thread score, the false alarm ratio (FAR), the Heidke skill score (HSS), the true skill statistic (TSS) and the accuracy, which were 
used in our evaluations, are determined as follows \citep{han65, mur93, bar09, kub17},
\begin{eqnarray}
POD & = & \frac{TP}{TP + FN}, \\
CSI  & = & \frac{TP}{TP + FP + FN}, \\
FAR & = & \frac{FP}{TP + FP}, \\
HSS & = & \frac{2 (TP \cdot TN - FP \cdot FN)}{(TP + FN)(FN + TN) + (TP + FP)(FP + TN)}, \\
TSS & = & \frac{TP}{TP + FN} - \frac{FP}{FP + TN}, \\
Accuracy & = & \frac{TP + TN}{TP + FP + FN + TN}
\end{eqnarray}

\acknowledgments
This work is supported by KAKENHI grant Number JP15K17620, JP18H04451 and JST CREST. The data used here are courtesy of NASA/SDO and 
the HMI \& AIA science teams, as well as the Geostationary Satellite System (GOES) team.\\

\clearpage
%
%


\clearpage

%
%
%
\end{document}